\documentclass[a4paper,12pt]{article}
\usepackage{amsmath}
\usepackage{amssymb}
\usepackage{graphicx}
\usepackage[usenames]{color}
\usepackage{latexsym}
\usepackage{mathrsfs}
\def\al{\alpha}

\def\ga{\gamma} 
\def\ep{\epsilon}
\def\lam{\lambda}

\def\th{\theta}
\def\sig{\sigma}
\def\Th{\Theta}
\def\alp{{\alpha'}}

\def\sigp{{\sigma'}}
\def\parn              {  \par\noindent }

 \def\calH{{\cal H}} 
  \def\calL{{\cal L}}
  
\def\calP{{\cal P}}  \def\calR{{\cal R}}
\def\calS{{\cal S}} \def\calT{{\cal T}} 
  \def\calX{{\cal X}}
 
\def\matrixii#1#2#3#4            {  \left(\begin{array}{cc}#1&#2\\#3&#4
                                       \end{array}\right) }
\def\vecii#1#2      {  \left(\begin{array}{c}#1\\#2\end{array}\right)  }
\def\del        {  \partial }
\def\half       {  {1\over 2}  }

\def\abs#1      {  \vert #1 \vert  }
\def\ie         {  {{\it i.e.}\  }    }

\def\comma          {\, ,}
\def\period         {\, .}
\def\lsim      {\lower .65ex \hbox{\ $\stackrel{<}{\sim}$\ } }
\def\gsim      {\lower .65ex \hbox{\ $\stackrel{>}{\sim}$\ } }

\def\com#1#2{{ \left[#1, #2\right] } }
\def\acom#1#2{{ \left\{ #1,#2\right\} } }
\def\nn               {  \nonumber  }
\newcommand{\nullify}[1]{}
\def\atil{\tilde{a}}

\def\btil{\tilde{b}}

\def\Qtil{\widetilde{Q}}

\def\Stil{\widetilde{S}}
\def\util{\tilde{u}}

\def\altil{\tilde{\alpha}}

\def\deltil{\tilde{\del}}

\def\ttil{\tilde{t}}
\def\omegatil{\tilde{\omega}}

\def\lamtil{\tilde{\lambda}}

\def\sigtil{\tilde{\sigma}}

\def\Pitil{\widetilde{\Pi}}

\def\adot{{\dot{a}}}

\def\muhat{\hat{\mu}}

\def\pp{{p^+}}

\def\sp{\sigma_+}
\def\sm{\sigma_-}

\def\pplusup{p^+}
\def\pminusup{p^-}

\def\Xp{X^+}
\def\Xm{X^-}

\def\lamp{\lambda^+}
\def\lamm{\lambda^-}
\def\lamtilp{\tilde{\lambda}^+}
\def\lamtilm{\tilde{\lambda}^-}

\def\piplus#1{\pi^{+#1}}

\def\Omp{\Omega^+}

\def\calXpl{\calX^+_L}
\def\calXpr{\calX^+_R}

\def\Pcom#1#2{\left\{#1,#2\right\}_P}
\def\Dcom#1#2{\left\{#1, #2\right\}_D}

\def\deltassp{\delta(\sig -\sigp)}
\def\deltapssp{\delta'(\sig-\sigp)}

\def\varth{\vartheta}

\def\ls{\ell_s}

\def\circbox{\hbox{$\scriptscriptstyle\circ$}}
\def\timesbox{\hbox{$\scriptscriptstyle\times$}}
\def\anc{{{\lower 1ex \circbox} \atop {\raise 1.5ex \circbox}}}
\def\ant{ {{\lower 1ex  \timesbox} \atop {\raise 1.5ex  \timesbox}}}
\def\ap{{\raise 0.2ex \hbox{$\scriptstyle \odot$}}}

\def\papertitlepage{\baselineskip 3.5ex \thispagestyle{empty}}
\def\Title#1{\baselineskip 1cm \vspace{1.5cm}\begin{center}
 {\Large\bf #1} \end{center}
\vspace{0.5cm}}
\def\Authors#1{\begin{center} {\it #1} \end{center}}
\def\Abstract{\vspace{1.0cm}\begin{center} {\large\bf Abstract}
           \end{center} \par\bigskip}
\def\Komabanumber#1#2#3{\hfill \begin{minipage}{4.2cm} UT-Komaba #1
              \parn #2
              \parn #3 \end{minipage}}
\renewcommand{\thefootnote}{\fnsymbol{footnote}}
\renewenvironment{thebibliography}{\pagebreak[3]\par\vspace{0.6em}
\begin{flushleft}{\large \bf References}\end{flushleft}
\vspace{-1.0em}

\begin{enumerate}\if@twocolumn\baselineskip=0.6em\itemsep -0.2em
\else\itemsep -0.2em\fi\labelsep 0.1em}{\end{enumerate} }

\begin{document}
\papertitlepage
\vspace*{0cm}
\Komabanumber{08-11}{ArXiv:yymm.nnnn}{July, 2008}
\Title{Superstring in the pp-wave background  \\
with RR  flux
as a conformal field theory\footnote[1]{Invited talk given  at the conference 
 ``30 Years of Mathematical Methods in High Energy 
Physics", held at RIMS, Kyoto University, March 2008.  To appear in the Proceedings.  }}
\Authors{{\sc Yoichi Kazama\footnote[2]{kazama@hep1.c.u-tokyo.ac.jp}
\\ }
\vskip 3ex
 Institute of Physics, University of Tokyo, \\
 Komaba, Meguro-ku, Tokyo 153-8902 Japan \\
  }
\baselineskip .7cm
\numberwithin{equation}{section}
\numberwithin{figure}{section}
\numberwithin{table}{section}
\parskip=0.9ex
\Abstract
We provide a concise description of our recent work on the exact
conformal field theory ( CFT)  formulation 
of the superstring in the pp-wave background with Ramond-Ramond (RR) flux, 
using the Green-Schwarz formalism in the semi-light-cone conformal gauge. 
Due to the presence  of the RR flux, the left- and the right-moving degrees of freedom 
on the worldsheet are  inherently coupled and this makes the canonical analysis  intractable even at the classical level.  To overcome this difficulty, we develop a phase-space 
formulation which does not  make use of the equations of motion,  and  construct two 
independent sets of Virasoro generators classically.  Upon quantization,  due to 
the presence of the RR flux,  the Virasoro algebra  generically  develops  quantum 
operator anomalies and only with a judicious choice of  normal ordering prescription
they cancel between the bosonic and the fermionic  contributions.  With such  properly 
defined quantum Virasoro generators, one can construct the BRST operators
and demonstrate that the BRST cohomology analysis reproduces   the physical spectrum 
previously obtained  in the light-cone gauge. 
\baselineskip 3.5ex
\section{Introduction}
\renewcommand{\thefootnote}{\arabic{footnote}}
Since its inception about 10 years ago,  an impressive collection of ``evidence" has
 been accumulated for the AdS/CFT correspondence\cite{Maldacena:1997re,Gubser:1998bc,
Witten:1998qj}, one of the most profound structures  in string theory. 
On the CFT side,  extremely detailed analyses have recently become possible, based on the 
powerful  assumption of ``integrability"\cite{Minahan:2002ve, Beisert:2003tq,
Beisert:2003yb, Beisert:2004ry} as well as on the state of the art perturbative techniques\cite{Bern:2005iz, Bern:2006ew, Bern:2007ct}. On the string side, we  have also acquired many interesting results, but most of them
  so far are classical\cite{Gubser:2002tv, Mandal:2002fs, Bena:2003wd,
Tseytlin:2004xa}, \ie based on the classical 
solutions of either the supergravity or the  string sigma model. Understanding of the 
 stringy aspects has been  slow due to the difficulty of solving the string 
 theory in the highly curved background with a large RR flux. 

In any case, it is fair to say that, despite the tremendous efforts having been made, 
the understanding of the fundamental mechanism 
 of this remarkable correspondence remains as  the central problem. 

As we know, although the AdS/CFT correspondence contains  some aspects 
 of the conventional open/closed string duality, there are crucial differences:
\begin{itemize}
	\item It is a strong-weak duality, not the usual perturbative duality. 
\item  It is a holographic duality. It is not relating  the closed and the 
 open strings both  in the bulk. 
\item  In contrast to the usual open-closed duality, there is a regime where both 
 the open and the closed channel descriptions are dominated by  the massless multiplets only, 
for example  by  the  super-Yang-Mills multiplet and the supergravity multiplet. 
\end{itemize}

These aspects must all be related,  but let us focus here on the strong/weak nature of the 
duality.  The familiar fundamental relation expressing this property is 
\begin{align}
g^2_{YM}N= 4\pi g_s N &=  {R^4 \over {\al'}^2} \period \label{fundrel}
\end{align}
The first equality expresses the ususal perturbative relation  between 
 the open and the closed string couplings. It is the second equality which relates the 
weak string coupling to the strong sigma-model coupling and vice versa. 
Let us recall the origin of this relation.  The bosonic part of the type IIB supergravity 
 action is of the form 
\begin{align}
S &= 
\int d^{10}x \sqrt{-g} \left( g_s^{-2}l_s^{-8}
 \calR -{2 \over 5!}F_5^2\right) 
\end{align}
We know that this admits the D3-brane solution due to the balance between the 
curvature $\calR$ and the RR $F_5$ flux. In the near horizon limit, this is 
 expressed as 
\begin{align}
g_s^{-2} {1\over l_s^8 R^2} &\sim \left({N \over R^5}\right)^2  \comma 
\end{align}
where $R$ is the common radius of $AdS_5$ and $S^5$ spaces.  Upon rearrangement 
 this immediately gives (\ref{fundrel}). 
This clearly shows  that the presence  of  the large RR flux is indispensable  for the 
strong/weak duality. 
It is acting as a kind of ``anti-gravity" preventing  the spacetime from collapsing. 

Unfortunately, the treatment of the RR flux (and  the associated curved background)
 is precisely 
 the largest obstacle which has been hampering the progress on the string side of 
 the AdS/CFT correspondence . To this date, the only such background in which 
 the string theory has been solved, to a certain  extent, is  the pp-wave background\cite{Penrose:1976, Blau:2001ne,
Blau:2002dy}. 
By adopting the light-cone (LC) gauge in the Green-Schwarz formalism\cite{Green:1983wt, Green:1983sg, Metsaev:1998it}, the exact  LC energy 
spectrum  was found \cite{Metsaev:2001bj} to be that of free massive bosons and fermions of the form 
$E_n = \sqrt{\mu^2 + (n/\alp p^+)^2}$, where $\mu$ is the ``mass parameter" 
 characterizing the curvature and the RR flux and $p^+$ is the momentum in the 
light-cone direction. This fact was exploited  by Berenstein et al (BMN)\cite{Berenstein:2002jq} to initiate a detailed  comparison of  the string  spectrum and that
of the anomalous dimensions  of the corresponding gauge-invariant operators in super-Yang-Mills theory (see \cite{Plefka:2003nb, Sadri:2003pr} for reviews). 
The interactions among such stringy modes, however, have been understood 
 only partially. The general form of the 3-point vertex has been studied in the framework 
 of the light-cone string field theory\cite{Spradlin:2002ar, Spradlin:2002rv, Constable:2002hw, Dobashi:2002ar, Gomis:2002wi, Pankiewicz:2002tg, DiVecchia:2003yp, Pankiewicz:2003ap, Dobashi:2004nm, Spradlin:2003xc},  while  the higher point functions have not been
 obtained.  
Thus, there is still a lot to be learned  from  the string theory in this simple yet 
prototypical background in order to unravel, in particular, 
 the role of the large RR flux in the AdS/CFT correspondence.

Now one of the major factors  preventing further developments of this theory is 
the lack of conformal  invariance in the light-cone gauge treatment. 
As it is a string theory, it should be possible to formulate it  as a  {\it conformal field theory (CFT)}, which was so powerful  in the case of the flat background.  More explicitly, 
let us list some of the concrete motivations for developing   CFT description:
\begin{itemize}
	\item It would be extremely interesting to understand how a (left-right coupled) ``massive theory"  can be understood as a CFT, especially since the string  in the 
$AdS_5\times S^5$ also has such an inherent left-right coupling. 
\item Even at the classical level, the Virasoro algebra structure for the superstring 
 in the pp-wave background with RR flux has not been discussed\footnote{With
NSNS flux in  RNS formalism, see \cite{Chizaki:2007sc}. }. 
\item One wishes  to be able to compute the correlation functions using a worldsheet 
description. 
Once one can do this exactly in $\alp$, we will obtain the corresponding information 
 in the gauge theory to all loop orders. 
\item It should be useful in understanding the  modular invariance
property.  In the LC gauge, the  modular $S$ transformation 
($\tau \leftrightarrow \sigma$) alters the gauge condition itself and the modular 
 invariance is not seen directly)\cite{Bergman:2002hv, Takayanagi:2002pi}. 
\item It should serve as  a  step towards developing a fully covariant pure spinor formalism
\cite{Berkovits:2000fe, Berkovits:2002zv, Berkovits:2002vn}
 for the pp-wave background in operator formulation. 
\end{itemize}

In the rest of this article, we will give a concise description  of our recent work\cite{Kazama:2008as} on  such a CFT description of the  plane-wave string theory. 
 In section 2. we will begin with  the 
attempt at classical canonical analysis 
 in (conformally-invariant) semi-light-cone gauge\cite{Carlip:1986cy, Carlip:1986cz, Kallosh:1987vq} and  describe the difficulty  one encounters
 in such an analysis. 
To overcome this difficulty, we develop, in section 3, a phase space formulation without the use of equations of motion. In section 4. we will perform the quantization, construct 
 the quatnum Virasoro operators carefully  and show that the  physical spectrum 
 agrees with the 
one computed in the light-cone gauge. In section 5 we will discuss 
 remaining issues. 
\section{Attempt at canonical analysis of classical  superstring in pp-wave background in conformal gauge}
\subsection{Green-Schwarz Lagrangian in the semi-light-cone gauge}
We will employ the Green-Schwarz formulation developed by 
Metsaev\cite{Metsaev:2001bj} (see also \cite{Metsaev:2002re}). The basic fields are the 10 
string coordinates\footnote{We use the convention $X^\pm\equiv 
 {1\over \sqrt{2}}(X^9\pm X^0)$. }
$X^\mu = (X^+, X^-, X^I)$,  with $I=1\sim 8$,  and the two sets of 
16-component Majorana spinors $\th^A_\al = (\theta^A_a, \theta^A_\adot)$, 
with $A=1,2$, where $a$ and $\adot$ respectively denote  the $SO(8)$ chiral and anti-chiral 
 indices. The local $\kappa$-symmetry is fixed by the semi-light-cone (SLC) gauge condition
$\ga^+\theta^A =0$, where $\ga^+ \equiv  \half (\ga^9 + \ga^0)$. 
This is equivalent to $\theta^A_\adot = 0$ and hence  only the $SO(8)$ chiral 
 components $\th^A_a$ will be present. 

With some slight redefinitions and simplifications, the Lagrangian in the SLC gauge 
constructed   by Metsaev\cite{Metsaev:2001bj}  can be cast into the form
\begin{align}
\calL &= \calL_{kin} + \calL_{WZ} \comma \\
\calL_{kin} &= -{T\over 2}\sqrt{-g} g^{ij} 
\left( 2\del_i \Xp \del_j \Xm + \del_i X^I \del_j X^I  -\mu^2 X_I^2 \del_i \Xp \del_j \Xp 
\right) \\
& +i\sqrt{2} T \sqrt{-g} g^{ij} \left(\del_i  \Xp ( \th^1 \del_j \th^1 
+\th^2 \del_j \th^2)+2\mu \del_i \Xp \del_j \Xp  \th^1
 \th^2 \right)\comma  \label{Lkin2} \\
\calL_{WZ} &= -i\sqrt{2} T\ep^{ij} \del_i \Xp( \th^1  \del_j \th^1 
-\th^2 \del_j \th^2)\comma 
\end{align}
where  $\calL_{kin}$ and $\calL_{WZ}$  are, respectively,  the kinetic and 
 the Wess-Zumino part, 
 $T= 1/2\pi \al' $ is the string tension and $\mu$ represents the strengh of the RR flux. 
The worldsheet coordinates are denoted by $\xi^i=(\xi^0, \xi^1) = (t, \sigma)$. 
After generating the Virasoro constraints by varying $g_{ij}$, we take the 
 conformal gauge $g_{ij} =\eta_{ij}={\rm diag}\, (-1, +1)$. Note that couplings  to the curved geometry and to the RR flux give  quartic  interactions in addition to the cubic interactions, which are   present 
even for $\mu=0$. In contrast, in the light-cone gauge 
these non-linearities  disappear as one can set   $\del_i X^+ \del^i X^+ $ to  a constant. 
\subsection{Equations of motion and their solutions }
Despite the non-linearities  of the Lagrangian mentioned above, it is possible to 
obtain the general solutions of the 
equations of motion  exactly. 

First,  the equation of motion for $X^+$ is easily seen to be 
$\del_i \del^i X^+ =0$. So  $X^+$ is a free field having the following 
 mode expansion.  (We will use calligraphic letters to denote the fields satisfying the equation 
 of motion.)
\begin{align}
\calX^+ (\sp, \sm)&= \calX^+_L(\sp) + \calX^+_R(\sm)\comma  
\qquad \sigma_\pm \equiv t\pm \sigma \comma \\
\calXpl (\sp)&= {x^+ \over 2} + \ls^2 \pp \sp + i\ls \sum_{n \ne 0}{1\over n} 
\altil^+_n e^{-in\sp}\comma \\
\calXpr (\sm)&= {x^+ \over 2} + \ls^2 \pp \sm + i\ls \sum_{n \ne 0}{1\over n} 
\al^+_n e^{-in\sm}\period 
\end{align}

Next consider the equation of motion for the transverse coordinate  $X^I$. It involves 
 the $\calX^+_{R,L}$ fields and reads 
\begin{align}
\del_+\del_- X^I + \mu^2 (\del_+\calXpl \del_- \calXpr) X^I =0 \period 
\label{eqmxi}
\end{align}
This can be solved by introducing new variables $\rho_\pm$  (effectively adopting the light-cone  frame)
\begin{align}
(\sp, \sm) &\rightarrow (\rho_+, \rho_-) \equiv (\calXpl(\sp), \calXpr(\sm)) \comma \\
\deltil_\pm &\equiv {\del \over \del \rho_\pm} = (\del_\pm \rho_\pm)^{-1}\del_\pm \period 
\end{align}
Then the equation (\ref{eqmxi}) simplifies to
\begin{align}
\deltil_+ \deltil_- X^I + \mu^2 X^I =0 \period 
\end{align}
Let us further form the following  field-dependent ``light-cone  coordinates"
\begin{align}
\ttil &\equiv {1\over 2\ls^2\pp} (\rho_++ \rho_-) \comma 
\qquad \sigtil \equiv {1\over 2\ls^2\pp} (\rho_+ - \rho_-) \period
\end{align}
Then, the equation becomes that of a  free massive field 
\begin{align}
\left( {\del^2 \over \del \ttil^2} - {\del^2 \over \del \sigtil^2} 
\right) X^I + M^2 X^I =0 \comma 
\end{align}
where the dimensionless ``mass" $M$ is defined  as $M \equiv  2\ls^2 \pp \mu$. 
The general solution $2\pi$ periodic in $\sigma$ can be written as\footnote{This solution 
was obtained in the appendix of \cite{Chizaki:2006pq} .}
\begin{align}
\calX^I &= \sum_n (a^I_n u_n + \atil^I_n \util_n ) \comma  \\
u_n &= e^{-i (\omega_n \ttil +n \sigtil)} =  e^{-i (\lamp_n \calXpr + \lamm_n \calXpl)} \comma \label{defu} \\
\util_n &=e^{-i (\omegatil_n \ttil -n\sigtil)} =  e^{-i (\lamtilm_n \calXpr + \lamtilp_n \calXpl)} \period 
\end{align}
Here  $a^I_n$ and $\atil^I_n$ are constant coefficients  and $\lam^\pm_n$ etc. are given by 
\begin{align}
\lam^\pm_n &= {1\over 2\ls^2 \pp} (\omega_n \pm n)\comma \quad 
\lamtil^\pm_n = {1\over 2\ls^2 \pp} (\omegatil_n \pm n) \comma  \nn\\
\omega_n &= \omegatil_n = {n \over |n|} \sqrt{n^2 + M^2} \qquad \mbox{for $n \ne 0$} 
\comma \nn\\
\omega_0 &= -\omegatil_0 = M \period \nn
\end{align}
$u_n$ and $\util_n$ consist of product of left- {\it and} right-going functions. 

Next, let us consider the equations of motion for the fermions $\th^A$. 
In terms of the LC-frame coordinates $\rho_\pm$, they read
\begin{align}
\deltil_+ \th^1&= -  \mu  \th^2  \comma \qquad
\deltil_-\th^2=  \mu \th^1  \period 
\end{align}
Combining them we get $\deltil_+ \deltil_- \th^A + \mu^2 \th^A =0$, which is 
the same equation satisfied by $X^I$. Therefore the general solution can be written 
 in terms of the functions $u_n$ and $\util_n$ as 
\begin{align}
\vartheta^A &= \sum_n ( b^A_n u_n + \btil_n^A \util_n)  \\
\mbox{with }\quad  \mu b^2_n &= i \lamp_n b^1_n \comma \qquad \mu \btil^2_n = i \lamtilm_n \btil^1_n   \period 
\end{align} 

Finally, look at the equation of motion for $X^-$. It takes the form 
\begin{align}
\deltil_+\deltil_- X^- &= \mu^2 \calX^I (\deltil_+ +\deltil_-) \calX^I 
+ i\sqrt{2}\mu (\varth^1  \deltil_+\varth^2 
- \varth^2 \deltil_-\varth^1  ) \period \nn
\end{align} 
Since the RHS consists of  known functions,  it can be easily solved for $X^-$ 
by inverting the Laplacian $\deltil_+\deltil_- $ 
 with some appropriate boundary condition. 

In this way one can  obtain  the general  classical solutions. But now we have an 
 apparent puzzle: How can we construct purely left- and right-going energy-momentum 
tensors out of the  basis functions $u_n(\sig_+, \sig_-)$ and $\util_n(\sig_+, \sig_-)$ ?
\subsection{Energy-momentum tensor}
To solve this puzzle, let us look at  the energy-momentum tensors $\calT_\pm$ obtained through  standard  procedure. In the  $\rho_\pm$ basis they read 
\begin{align}
{\calT_\pm  \over T}&= 
 (\del_\pm \rho_\pm)^2 \biggl[ \half \deltil_\pm  \Xp \deltil_\pm \Xm +  {1\over 4} (\deltil_\pm  X_I)^2  
 -{i \over \sqrt{2}} \deltil_\pm \Xp (\th^1 \deltil_\pm \th^1 + \th^2 \deltil_\pm \th^2)\nn\\
& -{1\over 4} (\deltil_\pm \Xp)^2 ( \mu^2 X_I^2 + 4\sqrt{2}  i \mu \th^1\th^2)
\biggr] \period
\end{align}
 We want to see if  $\calT_\pm$ are functions  of $\sig_\pm$, with the use of equations
 of motion.  Focus on $\calT_+$. Using 
$\deltil_+ \calX^+=1$ and $\mu \varth^2=-\deltil_+\varth^1$,  it  reduces to 
\begin{align}
\calT_+  &={T\over 2} (\del_+ \rho_+)^2 \biggl[  \deltil_+ \calX^- 
+  \half \left(  (\deltil_+ \calX_I)^2  -\mu^2\calX_I^2\right) 
  -i \sqrt{2}  (\varth^2 \deltil_+ \varth^2 - \varth^1 \deltil_+ \varth^1)\biggr] 
\period \label{calTplus}
\end{align}
This can be simplified drastically  upon expressing  $ \deltil_+ \calX^- $
 in terms of the other fields. 
Using various equations of motion, 
 the once-integrated equation  for $X^-$ can be written as 
\begin{align}
\deltil_+ \calX^-  &=  -\half \left(  (\deltil_+ \calX_I)^2  -\mu^2\calX_I^2\right) 
  +i \sqrt{2}  (\varth^2 \deltil_+ \varth^2 - \varth^1 \deltil_+ \varth^1)
+  f_+(\sp) \period
\end{align}
where $f_+(\sig_+)$ is an arbitrary function of $\sig_+$ produced through 
integration process. 
Subsitituting this into (\ref{calTplus}), we see that all the terms  containing physical fields $\calX^I$ and $\vartheta^A$ cancel, except for $f_+(\sp)$,  and 
$\calT_+$ collapses to an exceedingly compact expression
\begin{align}
\calT_+  &={T\over 2} (\del_+ \calX^+_L)^2 f_+(\sp) \period 
\end{align}
So we have a very peculiar situation. Although 
$\calT_+$ is indeed a function only of $\sp$,  $f_+(\sp)$ cannot be made  out 
 of  {\it local }  products of physical fields $\calX^I(\sp,\sm), \vartheta^A(\sp,\sm)$. 
Dependence on these fields must be through their integrals, \ie through $\sp$-independent modes $a^I_n, \atil^I_n,b^A_n, \btil^A_n$. 
Another notable feature is that  obviously the above form is not smoothly connected to 
 $\mu=0$ flat space case:  No matter how small $\mu$ is, as long as it is non-zero the classical 
 solutions are connected through this quantity and hence the energy-momentum tensors 
 take non-flat forms. 

How should  $f_\pm(\sig_\pm)$ be fixed ?  The requirement is that 
it must be determined so that the correct canonical  equal time commutation relations 
 are realized among the fields. 
To examine this, we now turn to the Poisson-Dirac  brackets for the fields and the modes. 
\subsection{Poisson-Dirac  brackets for the fields and the modes}
The bosonic momenta are given by 
\begin{align}
P^+ &= T \del_0 X^+ \comma \\
P^- &= T \bigl[\del_0 X^- - \del_0 X^+ ( \mu^2 X_I^2 + 4\sqrt{2} i \mu \th^1\th^2) 
 -2\sqrt{2} i (\th^1\del_+ \th^1 +\th^2\del_+\th^2)\bigr] \comma \\
P^I &= T \del_0 X^I   \comma 
\end{align}
while the fermionic momenta take the form
\begin{align}
p^1 &=i\sqrt{2}T (\del_0X^+ -\del_1 X^+) \th^1 
= i\piplus{1}\th^1 \comma \\
p^2 &= i\sqrt{2}T  (\del_0X^++\del_1 X^+) \th^2 
= i\piplus{2} \th^2 \period
\end{align}
Here the quantity $\pi^\pm$ are defined by 
\begin{align}
\piplus{1}&\equiv \sqrt{2} (P^+-T\del_1 X^+)
\comma \qquad \piplus{2} \equiv \sqrt{2} (P^++T\del_1 X^+) \period 
\end{align}
Of course the above definitions of the fermionic momenta should be regarded as 
 primary constraints 
\begin{align}
d^A &\equiv p^A - i\piplus{A}\th^A =0 \period 
\end{align}
We define the Poisson brackets as
\begin{align}
\Pcom{X^I(\sig,t)}{P^J(\sigp,t)} &= \delta^{IJ} \deltassp \comma 
\label{pcomXIPJ} \\
\Pcom{X^\pm(\sig, t)}{P^\mp(\sigp, t)} &= \deltassp \comma
\label{pcomXpmPmp} \\
\Pcom{\th^A_a(\sig,t)}{p^B_b(\sigp,t)} &= -\delta^{AB} \delta_{ab}
\deltassp \comma \\
\mbox{rest} &=0 \period  
\end{align}
Under this bracket, the fermionic constraints $d^A_a$ form the second class algebra
\begin{align}
\Pcom{d^A_a(\sig,t)}{d^B_b(\sigp,t)} &= 2 i \delta^{AB} \delta_{ab} 
\piplus{A}(\sig,t) \deltassp\period 
\end{align}
Thus  we introduce the Dirac bracket in the standard  way. 
Then, $\theta^A$ become self-conjugate:
\begin{align}
\Dcom{\th^A_a(\sig,t)}{\th^B_b(\sigp,t)} &= {i \delta^{AB} \delta_{ab}
\over 2 \piplus{A}(\sig,t)}\deltassp \period \nn
\end{align}
It is convenient to define the new field $\Th^A_a$ by 
\begin{align}
\Th^A_a &\equiv \sqrt{2\pi^{+ A}}\, \theta^A_a \period
\end{align}
It enjoys the canonical bracket relation with itself of the form 
\begin{align}
\Dcom{\Th^A_a(\sig,t)}{\Th^B_b(\sigp,t)} &= i\delta^{AB} \delta_{ab}
\deltassp  \period
\end{align}
In fact  it is easy to check that the set $\{ X^\mu, P^\mu, \Th^A_a\}$ satisfy the canonical (anti)-commutation
 relations under the Dirac bracket. 

The next step is to find the commutation relations among the modes so that the fields satisfy 
 the canonical Poisson-Diract bracket relations at equal $t$. Here we encounter a grave difficulty. 
Let us recall the solution for the  transverse coordinate $\calX^I$. It  is given in terms of
 the functions $u_n$ and $\util_n$ as 
\begin{align}
\calX^I &= \sum_n (a^I_n u_n + \atil^I_n \util_n ) \comma \\
u_n &= e^{-i (\omega_n \ttil +n \sigtil)} =  e^{-i (\lamp_n \calXpr(\sm)
 + \lamm_n \calXpl(\sp))} \comma \label{defu} \\
\util_n &=e^{-i (\omegatil_n \ttil -n\sigtil)} =  e^{-i (\lamtilm_n \calXpr(\sm)
 + \lamtilp_n \calXpl(\sp))} \period 
\end{align}
To extract $a^I_n$ and $\atil^I_n$, one needs completeness relations for the functions
 $u_n$ and $\util_n$ at ``equal time".  Evidently, it is easy for the equal  $\ttil$ slice, 
 just as in the LC gauge, but extremely hard for the  equal $t$ slice of our interest. 
Formal expression  can be derived but it depends on the modes of $\calXpr, \calXpl$
 in an intractably complicated  way.  (The reason for this is that a 
non-trivial field-dependent conformal transformation is 
 involved between the symplectic structures in the canonical $(t,\sig)$ basis and the 
$(\ttil, \sigtil)$ basis.)  So at this point  the canonical analysis has to be abandoned. 
\section{Phase space formulation without the use of equations of motion}
\subsection{Basic observation  }
Fortunately, the difficulty encountered   in the canonical analysis  described above 
can be overcome by 
 the use of the phase-space formulation.  The basic observation is as follows. 

In  ordinary field theories, the knowledge of the Poisson(-Dirac) brackets 
at equal $t$  is not enough to describe the dynamics which relates  different $t$. 
This is precisely the reason why we follow the canonical procedure. Namely, 
one first try to find the brackets for $t$-independent modes and then 
compute the brackets for fields at arbitrary  (unequal)  times. 

But  the situation is different for a string theory in the conformally invariant gauge. 
It is a type of theory in which the Hamiltonian $H$ is a member of 
 the generators of a large symmetry algebra, called the  ``spectrum generating algebra". 
In such a case, the representation theory of the algebra should know about  the spectrum and
 the dynamics: The analysis of the Virasoro constraints  give the information of the spectrum 
 and by constructing the primary fields one should be able to compute the correlation 
 functions which carry the dynamical information. In such a case, one may 
 use the phase space formulation, where the equations of motion are not needed explicitly
 and only the equal-time brackets  should be sufficient  to develop the representation theory. 
\subsection{Classical Viraosoro algebra in the phase space formulation}
Let us now construct the Virasoro generators in terms of the phase space variables 
at the classical level. 
To simplify the description, let us introduce  dimensionless fields $A, B, \calS, \Pitil, \Pi $ and a constant $\muhat$ as
\begin{align}
A &= \sqrt{2\pi T}X \comma \quad B = \sqrt{{2\pi \over T}} P 
  \comma \quad \calS = \sqrt{2\pi}\Theta \comma   \\
\quad \Pitil &= {1\over \sqrt{2}} (B +\del_1 A) \comma \quad \Pi 
 = {1\over \sqrt{2}} (B -\del_1 A)\comma \quad
 \muhat = {\mu\over \sqrt{2\pi T}}  \period 
\end{align}
It is useful to remember that the fields $(\{\Pitil^\star\},  \calS^2)$ 
  are ``left-moving", while $(\{\Pi^\star\}, \calS^1)$ are``right-moving". 
Then, the two sets of energy-momentum tensors can be written as
\begin{align}
\calT_+ &= \half (\calH + \calP)
= {1\over 2\pi} \biggl(\Pitil^+\Pitil^- + \half \Pitil_I^2 -{i \over 2} \calS^2 \del_1 \calS^2\nn\\
& \qquad + {\muhat^2 \over 2} A_I^2 \Pitil^+ \Pi^+ + {i\muhat \over \sqrt{2}}
\sqrt{\Pitil^+\Pi^+} \calS^1 \calS^2 \biggr) \comma \\
\calT_- &= \half (\calH - \calP) ={1\over 2\pi} \biggl( \Pi^+\Pi^- + \half \Pi_I^2 +{i \over 2} \calS^1 \del_1 \calS^1 \nn\\
& \qquad + {\muhat^2 \over 2} A_I^2 \Pitil^+ \Pi^+ + {i\muhat \over \sqrt{2}}
\sqrt{\Pitil^+\Pi^+} \calS^1 \calS^2 \biggr)\period 
\end{align}
where $\calH$ is the Hamiltonian density and $\calP$ is the momentum density. 
Although $\calT_+$ and $\calT_-$ do not manifestly commute with each other, 
after some careful computations, we verify the two independent sets of closed algebras
\begin{align}
  &\Dcom{\calT_\pm(\sig,t)}{\calT_\pm(\sigp,t)}  
 = \pm 2 \calT_\pm(\sig,t)\deltapssp \pm 
\del_1 \calT_\pm(\sig,t) \deltassp\comma  \label{classvir} \\
&\Dcom{\calT_\pm(\sig,t)}{\calT_\mp(\sigp,t)} =0 \period 
\end{align}
These relations contain the information of the time-development. In particular, 
\begin{align}
\del_0\calH &= \Pcom{\calH}{H} = \del_1 \calP \comma \qquad 
\del_0\calP = \Pcom{\calP}{H} = \del_1 \calH \period 
\end{align}
where $H  \equiv \int d\sig \calH$ is the Hamiltonian. 
By adding and subtracting these equations,  we  readily get 
$\del_\pm \calT_\pm =0$, which means
\begin{align}
 \calT_\pm= \calT_\pm(\sigma_\pm) \period \label{holomvir}
\end{align}
From (\ref{classvir})  and  (\ref{holomvir})  we learn  
that the modes of $\calT_\pm$ form the (classical) Virasoro  algebra:
\begin{align}
\calT_\pm &= {1\over 2\pi} \sum_n T^\pm_n e^{-in\sig_\pm}   \comma  \\
\Dcom{T^\pm_m}{T^\pm_n} &= {1\over i} (m-n) T^\pm_{m+n} 
\comma \qquad \Dcom{T^\pm_m}{T^\mp_n} =0 \period 
\end{align}
Note that to extract the modes we only need the information at $t=0$, namely 
\begin{align}
T^\pm_n &= \int_0^{2\pi}  d\sig e^{\pm in\sig} \calT_{\pm}(\sig, t=0) \period
\end{align}
Thus, despite the left-right coupling, we have two independent sets of 
 classical  Virasoro algebras for any value of $\mu$. It is interesting to observe that 
 viewed  as exactly marginal deformations  from the flat space case they are 
somewhat unusual since the $\mu$-dependent terms are not  primary 
 with respect to the $\mu=0$ (flat space) theory. 
\section{Quantization, quantum Virasoro algebra,  and the physical spectrum}
\subsection{Quantization}
The quantization of the  basic fields is done  by replacing 
 the Poisson-Dirac brackets  by  quantum 
 brackets  in the usual way at $t=0$. 
For example,  
\begin{align}
\Dcom{\Pitil^+(\sig)}{\Pitil^-(\sigp)} &=  2\pi \deltapssp  \ \ 
\Rightarrow \ \  \com{\Pitil^+(\sig)}{\Pitil^-(\sigp)} &=  2\pi i \deltapssp 
\period 
\end{align}
As for the mode expansion at $t=0$, we will adopt the following convention 
\begin{align}
\phi(\sig)&= \sum_n  \phi_n e^{-in\sig} \period 
\end{align}
Then,  the commutator of the modes take the form such as 
\begin{align}
\Pitil^\pm (\sig) &= \sum_m \Pitil^\pm_m e^{-im\sig}  \comma \qquad 
\com{\Pitil^\pm_m}{\Pitil^\mp_n} = m \delta_{m+n, 0} \comma \qquad etc. 
\end{align}
\subsection{Quantum Virasoro algebra }
Now we come to the construction of the quantum Virasoro operators. For this purpose, 
 we must (i) find an appropriate normal-ordering,  (ii) make sure that the central charges
 add up to $26$ and (iii) add quantum corrections, if necessary. 

Finding the correct normal-ordering turned out to be quite non-trivial. 
To describe and check the appropriate scheme we found, it is convenient to introduce 
the operators $\calL_\pm(\sig)$, which satisfy  the same form of the Virasoro
 algebra (as opposed to $\calT_\pm$ which satisfied the relations (\ref{classvir}) with 
different signs) 
\begin{align}
\calL_\pm (\sig) &\equiv  \pm \calT_\pm (\sig) = {1\over 2\pi} 
\sum_n L^\pm_n e^{-in\sig} \period
\end{align}
Now to define $L^\pm_n$ quantum-mechanically, we adopt the 
``phase-space  normal-ordering"
where $B^\star_n (n \ge 0), A^\star_n (n \ge 1), S^A_{a,n}(n \ge 1)$ are 
regarded as annihiliation operators. The crucial feature of this normal-ordering is that 
with such a prescription the 
quantum operator anomalies produced by the double contractions  between 
 the non-linear terms cancel exacly between the bosonic and fermionic contributions\footnote{
This in particular means that in the plane-wave background {\it with RR flux},
  the Virasoro operators  made out of bosonic fields alone  cannot close 
quantum-mechanically in a consistent way. }. 
These anomalous contributions 
 are of the form (the subscript $B(F)$ stands for bosonic(fermionic)) 
\begin{align}
C_B &= {1\over (2\pi)^2} \left(\com{\half \Pitil_I^2(\sig)}{{\muhat^2 \over 2}\Pitil^+\Pi^+ A_I^2(\sigp)}
-(\sig \leftrightarrow \sigp) \right) \comma \\
C_F &=  {1\over (2\pi)^2}  \com{{-i\muhat\over \sqrt{2}}\sqrt{\Pitil^+\Pi^+}
S^1S^2(\sig)}{{-i\muhat\over \sqrt{2}}\sqrt{\Pitil^+\Pi^+}S^1S^2(\sigp)} 
\period
\end{align}
Upon careful computation with appropriate  regularization  one finds 
\begin{align}
C_B &= -C_F = -{i \muhat^2 \over \pi} ( 2 \Pitil^+\Pi^+ \deltapssp
+ \del_\sig(\Pitil^+\Pi^+) \deltassp )  \period 
\end{align}
With this cancellation, one verifies  that $L_n^\pm$ form two  sets 
 of  quantum Virasoro algebra:
\begin{align}
\com{\calL_\pm(\sig)}{\calL_\pm(\sigp)} &= i \biggl( 2\calL_\pm(\sig) \deltapssp 
 + \del_\sig \calL_\pm(\sig) \deltassp \nn \\
&\qquad  -{1\over 24\pi} (14 \delta'''(\sig-\sigp) -2\deltapssp) \biggr) \comma 
\label{quantvir}\\
\com{\calL_+(\sig)}{\calL_-(\sigp)} &= 0 \period 
\end{align}

As is evident from the form of (\ref{quantvir}), the central charge is only 14, 
 10  from $X^\mu$ and 4 from $S^a$. 
This is the same as in the flat space case
 in the SLC gauge and the cure is known\cite{Berkovits:2004tw}. One needs to add the 
 quantum corrections $\Delta \calL_\pm$ of the following form:
\begin{align}
\calL_\pm &\longrightarrow \calL_\pm + \Delta \calL_\pm  \comma \\
\Delta \calL_+ &= -{1\over 2\pi} \del_\sig^2 \ln \Pitil^+ \comma \qquad 
\Delta \calL_- = {1\over 2\pi} \del_\sig^2 \ln \Pi^+ \period 
\end{align}
$\Delta \calL_\pm$ almost behave as  primary operators of dimension 2, except that they 
 provide the wanted  12 units of central charge. One can indeed verify 
\begin{align}
&\com{\calL_\pm (\sig)}{\Delta \calL_\pm (\sigp)} + 
\com{\Delta \calL_\pm (\sig)}{\calL_\pm (\sigp)} \nn\\
& = i \biggl( 2 \Delta \calL_\pm (\sig) \deltapssp + \del_\sig (\Delta \calL_\pm (\sig)) \deltassp
-{1\over 24\pi} 12 \delta'''(\sig-\sigp) \biggr) \comma 
\end{align}
so that with this addition we have the desired quantum Virasoro operators with 
 central charge $26$. 
\subsection{BRST formulation  and the physical spectrum}
Having constructed the quantum Virasoro operators, it is now straightforward to 
 construct the nilpotent BRST operators $Q$ and $\Qtil$ for the right- and the left-sector. 
$Q$, for instance, takes the familiar form
\begin{align}
Q &= \sum_n c_{-n}L^-_n -\half \sum_{m,n} (m-n) :c_{-m}c_{-n} b_{m+n}: 
\period 
\end{align}
Although the Virasoro generators  $L^-_n$ in this formula 
contain non-linear terms, the decoupling of the unphysical 
degrees of freedom, namely the  non-zero modes of 
$\Pitil^\pm, \Pi^\pm, b,c, \tilde{b}$ and $\tilde{c}$,  works in a simple way. 
This is because  one can easily prove  the equivalence 
$Q$-cohomology $\simeq$ $Q_{-1}$ cohomology, where $
Q_{-1} \equiv  -\Pi^+_0 \sum_{n\ne 0} \Pi^-_{-n} c_n$ is  exactly the same as in the 
free bosonic string\cite{Polchinski:1998rq}. 
Consequently, the physical states are the ones in the transverse space $\calH_T$ (\ie 
 without the non-zero modes above), 
 satisfying  the constraints $H=L^+_0 + L^-_0 = 0$,  $ P= L_0^+-L^-_0 =0$. 

After dropping  the  non-zero modes of $\Pitil^\pm, \Pi^\pm$, our Hamiltonian 
 becomes (with phase-space normal-ordering understood)
\begin{align}
H&= H_B + H_F \comma \\
H_B &= \alp \pplusup\pminusup + \half \sum (B^I_{-n} B^I_n 
+ \omega_n^2 A^I_{-n} A^I_{n})
\comma \\
H_F &= \half \sum (- n S^1_{-n} S^1_n +n S^2_{-n} S^2_n -iM S^1_{-n} S^2_n
+ iM S^2_{-n} S^1_n ) \comma 
\end{align}
where $\omega_n \equiv {n \over |n|} \sqrt{n^2+M^2}$. Evidently,  
$H_B$ describes  free massive bosonic excitations. As for $H_F$, we need to perform 
 diagonalization to see that it describes the corresponding   free massive fermionic excitations. To this end, construct  massive oscillators in terms of massless oscillators in 
 the following way. 
\begin{align}
\altil^I_n &\equiv  {1\over \sqrt{2}} (B^I_n -i \omega_n A^I_n) \comma \quad 
\al^I_n \equiv  {1\over \sqrt{2}} (B^I_{-n} -i \omega_n A^I_{-n})
\comma   \quad (n\ne 0) 
\\
\al_0^I & \equiv  {1\over \sqrt{2}} (B^I_0 -iM A^I_0) \comma \quad 
{\al_0^I}^\dagger \equiv  {1\over \sqrt{2}} (B^I_0 +iM A^I_0) \comma \\
\Stil_n &\equiv N(n) \left( S^2_n + i {M \over \Omp_n} S^1_n \right) 
\comma \quad 
S_n \equiv N(n) \left( S^1_{-n} - i {M \over \Omp_n} S^2_{-n} \right) 
\comma \quad (n\ne 0) \\
S_0 &\equiv {1\over \sqrt{2}} ( S^1_0 -i S^2_0) 
\comma \quad S_0^\dagger \equiv {1\over \sqrt{2}} ( S^1_0 +i S^2_0) \comma  \\
\Omp_n  &\equiv  \omega_n + n
 \comma \quad N(n) \equiv \sqrt{\Omp_n \over 2\omega_n} \period 
\end{align}
They satisfy the (anti-)commutation relations 
\begin{align}
\com{\altil^I_m}{\altil^J_n} &= \com{\al^I_m}{\al^J_n} 
=\omega_n \delta^{IJ} \delta_{m+n,0} \comma \qquad \com{\altil^I_m}{\al^J_n} =0 
\comma \\
\com{\al^I_0}{{\al^J_0}^\dagger} &= \delta^{IJ} M = \delta^{IJ}
\omega_0 \comma  \\
\acom{\Stil_m}{\Stil_n} &= \acom{S_m}{S_n} = \delta_{m+n,0} \comma \quad 
\acom{S_0}{S_0^\dagger} = 1 \comma \quad \acom{\Stil_m}{S_n} =0 
\period 
\end{align}
Now re-express $H$ in terms of  these oscillators and re-normal-order appropriately for  
 the new oscillators. Then one finds  that, due to supersymmetry, 
 the constants produced in this process cancel exactly and the Hamiltonian simplifies to 
\begin{align}
H &= \alp \pplusup\pminusup + {\al_0^I }^\dagger \al_0^I +
\sum_{n \ge 1} (\al^I_{-n} \al^I_n+\altil^I_{-n} \altil^I_n ) \nn\\
& + M S_0^\dagger S_0 +\sum_{n \ge 1}  \omega_n (S_n^\dagger S_n + \Stil_n^\dagger \Stil_n)  \period 
\end{align}
Setting $H=0$ and solving  for $-p^-=H_{lc}$, the light-cone Hamiltonian, 
we get
\begin{align}
H_{lc} &=  {1\over \alp \pp}\biggl( {\al_0^I }^\dagger \al_0^I +
\sum_{n \ge 1} (\al^I_{-n} \al^I_n+\altil^I_{-n} \altil^I_n ) \nn\\
&+ M S_0^\dagger S_0 +\sum_{n \ge 1} 
 \omega_n (S_n^\dagger S_n + \Stil_n^\dagger \Stil_n)  \biggr) \period
\end{align}
This  coincides with  the well-known light-cone Hamiltonian in the LC gauge\cite{Metsaev:2001bj}. Further, 
$P =0$ yields  the level-matching condition
\begin{align}
P &= \sum_{n \ge 1} \left( {n \over \omega_n} \altil^I_{-n} \altil^I_n 
+ n \Stil^\dagger_n \Stil_n \right) 
-\sum_{n \ge 1} \left( {n \over \omega_n} \al^I_{-n} \al^I_n 
+ n S^\dagger_n S_n \right)=0  \period
\end{align}
%
This shows that the physical spectrum of our conformal field theory is precisely 
the same as in the light-cone gauge. 
\section{Summary and remaining issues}
We have initiated the study of  the superstring in the pp-wave background with RR-flux 
 as an exact  conformal field theory in an operator formulation.
 Despite non-linearity, the equations of motion can 
be solved exactly. However,  the canonical 
 analysis based on these solutions meets difficulties:
Transverse fields are functions of both $\sig_+$ and $\sig_-$. Only their 
 modes  can appear in the Virasoro generators and it turned out to be 
extremely hard to find the commutation relations for  these modes. 

	 We pointed out that to overcome this difficulty   an alternative phase space formulation without  the use of equations of motion can be utilized. 
Despite the coupling between left- and right-going fields, two independent  sets of 
classical Virasoro generators are constructed. 
Fields can be quantized in a straightforward manner and in terms of them 
the quantum Virasoro generators are defined with appropriate normal-ordering and 
a quantum modification. They are checked to form correct Virasoro algebra. 
Also we have shown that they reproduce, via BRST formulation,  the correct physical spectrum previously obtained  in the light-cone  gauge.

Clearly, there are many remaining problems to be investigated. We must clarify  how the global symmetries are realized, including the supersymmetry. The most important 
 and a challenging task is the construction of the primary fields, at least for  the low  lying 
excitations. Once this is achieved, we should be able to compute the correlation functions
exactly in $\al'$.  It would also be of interest to construct the DDF operators  for
 all the excitations.  Study  of the modular invariance, which is awkward in the 
 light-cone gauge,  is another important problem. It is intriguing to see how the presence of 
 the large RR flux affects the nature of the open/closed duality.  The CFT we constructed 
 in the SLC gauge may be a starting point of  the operator formulation 
 of the fully covariant pure-spinor formalism, through the double spinor extension\cite{Aisaka:2005vn} of 
 the Green-Schwarz superstring.  Finally, we should study if our phase-space formalism 
 could  be applied to the case of superstring in the $AdS_5\times S^5$ background 
in a useful way. 
We hope to be able to report progress on these issues in a near future. 
\par\bigskip\noindent
{\large\bf Acknowledgment}\par\smallskip\noindent
It is a pleasure to thank the organizers of the conference 
``30 Years of Mathematical Methods in High  Energy Physics",  where this work 
 was presented,  for providing 
a stimulating atomosphere  and hospitalities. 
This work  is supported in part by the 
 Grant-in-Aid for Scientific Research (B) 
No.~12440060 and (C) No.~18540252 from the Japan 
 Ministry of Education, Culture, Sports,  Science and Technology.

\end{document}